\documentstyle{l-aa}

\begin{document}

\pagestyle{empty}

\thesaurus{19(16.01.1:19.10.1:19.15.1:19.43.1}

\title{Lithium Abundance in Binaries of the Hyades Open Cluster\thanks
{Based on observations obtained at Kitt Peak
National Observatory, Arizona, operated
by the Association of Universities for Research in
Astronomy, Inc., under contract
to the National Science Foundation, USA.}}

\author{ D. Barrado~y~Navascu\'es \inst{1,2,3} 
         \and J.R.~Stauffer \inst{2}}

\offprints{D.~Barrado~y~Navascu\'es.  dbarrado@cfa.harvard.edu\\
{\bf Astronomy And Astrophysics, (1996), 310, 879}}

\institute{Depto. Astrof\'{\i}sica, Facultad de F\'{\i}sicas,
           Universidad Complutense. 
           E-28040 Madrid, Spain.
        \and
	Harvard--Smithsonian Center for Astrophysics.
           60 Garden St.,
          Cambridge, MA 02138, USA.
        \and
           Fellow at the Real Colegio Complutense.
          26 Trowbridge St., 
          Cambridge, MA 02138, USA.}
                    
\date{Received date,  13 June 1995; accepted date, 25 September 1995}

\maketitle

\begin{abstract} 

     We have derived accurate and homogeneous Lithium 
     abundances in 49 main sequence binary systems 
belonging to the Hyades Open Cluster by using a deconvolution
method
to determine individual magnitudes and colors for the primary and 
secondary components of the binary. The input parameters
of the model are the observed Li equivalent width, the actual
distance  
to the binary, the integrated apparent magnitude and the
integrated 
colors of the binaries --BV(RI)$_{\rm K}$.

     We show that the general behavior is the same in  binaries
and in single stars
(Li is depleted faster in K stars than in G stars and there is a
deep dip for mid--F
stars). However, there is a larger scatter in the abundances of
binary 
systems than in single stars. Moreover, in general, binary
systems have
an overabundance, which is more conspicuous in close binaries. In
fact, there is a 
cut--off period, which can be estimated as P$_{\rm orb}$$\sim$9
d. 
This value is in
excellent agreement with the theoretical predition of Zahn
(1994). 

Table 1 is also available in electronic form
at the CDS via anonymous ftp 130.79.128.5

\end{abstract}

\section{Introduction}

The Hyades  open cluster is one of the essential  
laboratories in Astrophysics. It is an intermediate richness 
cluster with an age of 6--8$\times$10$^8$ yr (Mermilliod,
1981; 
Gilroy 1989). 
Due to its proximity (47.86 pc, Schwam 1991), it allows us
to obtain very accurate information about a variety of topics,
such  as color--absolute
magnitudes relations, the distance scale of the Galaxy and,
therefore, the
Universe, etc. It has also been essential in the determination of
the time
scale of a large range of phenomena.

In particular, the Hyades has been very important in the 
studies of the Lithium phenomenon. Li is destroyed 
near the bottom of the convective envelope in late spectral
type
Main Sequence (MS) stars. For this reason, Li is depleted with
age.
Since the late type members of the Hyades have known age and it
is possible
to obtain spectra with excellent signal--to--noise ratios (SNR),
the Hyades
has become a fundamental instrument to calibrate this dependence
(Boesgaard 1991). In fact, the 
Li depletion is also characterized by a dependence on stellar
mass (Cayrel et al.
1984),  the presence of a deep gap for mid--F stars older than
$\sim$100 Myr
(Boesgaard \& Tripicco
1986a), a  dependence on metallicity (Hobbs \& Duncan 1987), and
an apparent dependence on rotation at least for very young stars
(Marcy et al. 1985).
Lately, Garc\'{\i}a--L\'opez (1995) has introduced the expresion
Big Lithium Gap to describe the depletion which appears for stars between
G spectral type and the substellar limit, since Brown Dwarfs must not 
deplete Lithium (Rebolo et al. 1992).
The two first peculiarities were discovered in the Hyades. 
Soderblom et al. (1990) showed that some Hyades stars also have 
overabundances compared with the average value of stars of the
same color,
 and these excesses could be related with their 
binary nature. They also claim that 
 there is a real scatter --though smaller than
the scatter present in other younger 
clusters-- in the Li abundances for 
stars of the same color. The presence of this scatter was
confirmed
in an extensive study by Thorburn et al. (1993).
The role of rotation has been shown to be very important
both in open clusters, such as Pleiades (Soderblom et al. 1993a)
and in chromospherically active binaries (Pallavicini et al.
1992; Randich et al.
1993a,b; Fern\'andez--Figueroa et al. 1993), but it is not 
totally  clear
how rotation affects  the Li abundance.

 \begin{figure*}
 \vspace{18cm}
 \caption[ ]{ Normalized spectra around LiI6708 \AA{ } for 
(a) V471 Tau, (b) vA677, (c) J301, (d) J304, (e) vB 185
 (f) vB 117 and (g) J331.
Note the very broad
spectral lines in the   spectrum of V471~Tau and the
double--lined spectra of vB~185, J331 and vB~117.
 All primaries except V471~Tau, vA~677 and
J331 have very similar (B-V) color.
 }
 \end{figure*}

 Since  Li can be transported by different mechanisms  inside the
convective
envelope or below it, it provides an important way to 
discriminate which mechanism is more important
 and to look inside the internal structure 
of  low mass stars. These mechanisms,
 such as convection, meridional circulation (Charbonneau \&
Michaud 1988), 
microscopic diffusion (Michaud 1986), gravity waves 
(Garc\'{\i}a--L\'opez \& Spruit 1991; Montalban 1994),
 etc,  have  been used to explain one or more of 
the observed characteristics.
Among these proposed theoretical  explanations, the ones 
which involve rotation have the most potential to explain  
Li depletion in low mass stars. In particular,
 the mixing mechanisms due to the  angular momentum  loss
(Pinsonneault
at el. 1989) could  explain  not only the dependence
of the Li on age and stellar mass, but also the scatter for stars with
the same 
age and color  since the stars would have 
different initial angular momentum.
Lately, Zahn (1994) has found that 
meridional circulation, which itself appears due to the angular
momentum  transport between the radiative core and the convective
envelope,
 can explain the Li abundance of stars in different clusters and
in particular the high Li abundance of  close Hyades binaries.

Different studies have provided a large amount of data about the
stellar properties
of the binaries in the Hyades. 
At the present time, there are accurate data about the individual
distances
of the members (Schwan 1991; Stern 1994), extensive studies about
possible faint
companions of members (Mason et
al. 1993) and orbital elements of binaries (Griffin et al. 1988;
Stefanik
\& Latham 1992). With this information, we have attempted to 
verify whether the scatter in inferred Li abundance due
to the binarity is due only to the difficulty of the analysis,
or is due to other reasons, as can be 
the gravitational interaction between the
components. 
 
In this work, we present accurate Li abundances of binary
systems belonging
to the Hyades. Section 2 describes the sample and the original
sources of the
data. We show the method to correct the effect of the secondary star
on 
the colors and magnitudes in Section  3 and correction of the Li
equivalent
width --EW(Li)-- and the calculation of the abundances in Section
4. Section
5 contains the results and Section 6 the discussion and
interpretation.

\section{The program stars}

Our  sample is composed of Hyades binary systems for  which it
is possible to derive
 Li abundances.
The basic data for the Hyades binaries\footnote{Hereafter, we
will use the word {\it binary}
to refer to spectroscopic, photometric and visual binaries that
only 
have been separated by speackle interferometry, and the word {\it
single} to refer
 to real single and wide visual binaries.} 
used in this work have been
 selected essentially from Johnson \& Knuckles (1955)
--the photometry--, Stern (1994) --the distances-- and Thorburn
et al. (1993)
and this work
-the Li equivalent width, EW(Li). We have compiled  additional
information about 
EW(Li) from Duncan \& Jones (1983), Boesgaard \& Tripicco (1986a)
and Rebolo \& Beckman (1988) and about the photometry from
Mermilliod (1976). 
In total, we have analysed 52 stars belonging to 49 different
binaries.

     Absolute visual magnitudes for the combined components of
the binary 
(M$_{\rm V}^{\rm obs}$, column 6 of Tab. 1) were calculated by 
using the 
apparent visual magnitudes given by Johnson \& Knuckles (1955)
 and the distances given by Stern
(1994) when available. For vB~9 we used the average distance
modulus
 computed for our sample of binaries (${\rm
<m-M>=3.41^{+0.17}_{-0.19}}$), 
which is in excellent agreement with the value for the center of
the cluster 
(${\rm <m-M>=3.40\pm0.04}$) by  Schwan (1991), based on  145
stars with high
 accurate  proper motions.

	Another four binaries do not  have measured distances. However,
the difference between the magnitudes of both components has been measured
(see notes to Tab. 1). Therefore, it is possible to deconvolve the 
photometry with this information, and the solution is unique. Moreover,
photometric distances were calculated for these systems. These distances 
are listed  in Tab. 1.

     Finally, one of us (JRS) obtained  spectra of 7 additional
late spectral type binaries 
-(B-V)$\ge0.85^{\rm m}$- during October 1994. The observations were made
with the  4m telescope at 
 Kitt Peak National Observatory and the red, long camera echelle
 spectrograph.
The spectra were reduced with the routines provided within
IRAF\footnote{IRAF is distributed 
by National Optical Astronomy Observatories, which is operated by
the Association of Universities 
for Research in Astronomy, Inc., under contract to the National
Science Foundation, USA}, 
using standard procedures.
At our resolution (R$\sim$45000) and for rapid rotators,
the Li{\sc I} 6707.8 \AA{ } doublet is normally blended with
Fe{\sc I} 6707.4 \AA. We measured 
the total EW of the blend. Then, in order to obtain the EW(Li),
we either fitted two gaussian
 curves to the feature or assumed a EW(Fe) from the (B-V) color index
with the available data
for the Hyades. The agreement between our measured or estimated
EW(Fe{\sc I})s with those
values obtained with the empirical relation by Soderblom et al.
(1993a) is good.  
The observed stars, visual magnitudes, signal--to--noise
ratios, and EWs of different 
lines are given in Tab. 2. We show the normalized spectra in Fig.
1. Note the very broad
spectral lines in the   spectrum of V471~Tau (Fig. 1a) and the
double--lined spectra of vB~185, J331 and vB~117
(Fig. 1e, Fig. 1f and Fig. 1g). All primaries but V471~Tau, vA677 and
J331 have similar (B-V)$_{\rm primary}$ color. 
Casual inspection of these figures shows that the
 lines in
the double-lined  spectra (those systems having a
(1+$\alpha$) close to 2, see Section 4.2) appear dilute
because of the "extra" continuum light.


\footnotesize

\begin{table*}
\caption[  ]{Deconvolution for the Hyades binaries. The photometry is essentially from Johnson \&
Knuckles (1955) and the distances from Stern (1994). Spectral types for the primary and secondary components
were assigned from (B-V) color indices by using the Schmidt--Kaler's (1982) ZAMS. }
\begin{tabular}{llllllllclllcc}
\hline
Star               & \multicolumn{3}{c}{ (B-V)}      & Dist.          & \multicolumn{3}{c}{M$_{\rm V}$}&V$_{\rm obs}$& \multicolumn{3}{c}{Sp. Type}& Binarity\\
\cline{2-4} \cline{6-8} \cline{10-12}
                   &  Obs. & Prim. & Sec.   & (pc) & Obs.             & Prim.  & Sec.    &    & Obs. & Prim. & Sec. &        \\ 
     (1)           &  (2)  &  (3)  &  (4)   & (5)  &  (6)             & (7)    & (8)     & (9)& (10)    &(11)& (12) & (13)   \\
\hline
vB~1$^{(1)}$    & 0.566 & 0.559  &1.448 & 43.1             &4.228           & 4.241 & 9.028   &7.40&  --  & F9 V  & M1 V & VB     \\

V471 Tau$^{(2)}$& 0.85  &$\sim$0.993& --   & 59               & 5.86           & 5.35  & 9.75    &9.71&  --  & K2 V  & WD   & SB1    \\

vB~9            & 0.708 & --     & --   &(47.9)$^{\rm (b)}$&5.27$^{\rm (b)}$& --    & --      &8.67&  --  & G6 V  & --   & SB1 \\

vB~14           & 0.355 & 0.350  & 1.337& 38.3             & 2.814          & 2.822 & 8.141   &5.73& --   & F2 V  & K8 V & SB     \\

BD+23$^\circ635^{(3)}$& 1.09 & 1.043  &1.362 &(39.9)$^{\rm (a)}$&6.375           &6.573  & 8.322   &9.38& --   & K3.5 V& K9 V & SB2    \\

vB~162$^{(4)}$  & 0.705 & 0.666  & 0.746& 49.1             & 4.375          & 4.934 & 5.283   &7.83& --   & G4 V  & G9 V & SB2    \\

vB~22$^{(5)}$   & 0.771 & 0.713  & 0.945& 50.8             & 4.811          & 5.155 & 6.226   &8.34& G8 V & G6 V  & K3 V & SB2    \\

vB~23$^{(6)}$   & 0.679 & 0.654  & 0.714& 49.3             & 4.076          & 4.815 & 5.161   &7.54& --   & G3 V  & G6 V & SB2     \\

BD+22$^\circ669^{(7)}$& 0.99 & 0.933  & 1.167&(53.1)$^{\rm (a)}$& 5.859          & 6.190 & 7.310   &9.48& K0   & K2 V  & K5 V & SB2     \\

vB~29$^{(8)}$   & 0.561 & 0.520  & 0.670& 47.3             & 3.506          & 3.898 & 4.896   &6.88& F8 V & F8 V  & G5 V &  SB    \\

vB~34$^{(9)}$   & 0.457 & 0.435  & 0.461& 50.3             & 2.662          & 3.456 & 3.506   &6.17& F6 V & F5 V  & F6 V & SB2    \\

vB~38$^{(10)}$  & 0.320 & 0.296  & 0.764& 47.6             & 2.332          & 2.402 & 5.350   &5.72& Am   & F0 V  & G9 V & SB1   \\

vB~40$^{(11)}$  & 0.563 & 0.528  & 1.042& 41.9             & 3.879          & 3.974 & 6.567   &6.99& G9 V & F9 V  & K4 V &  VB/SB1 \\

vB~39$^{(12)}$  & 0.678 & 0.678  & --   & 38.2             & 4.950          & 4.950 & --      &7.86&  --  & G5 V  & --   & SB  \\

vB~42           & 0.759 & 0.750  & 1.492& 52.0             & 5.280          & 5.298 & 9.750   &8.86& G9 V & G8 V  & M2 V &  PhB  \\

vB~50           & 0.601 & 0.578  & 1.252& 45.9             & 4.311          & 4.361 & 7.668   &7.62& G1 V & G0 V  & K6 V  & SB     \\

vB~52$^{(13)}$  & 0.597 & 0.597  & --   & 43.9             & 4.588          & 4.588 &--       &7.80& G1 V & G0 V  & --    & SB?   \\

vB~140$^{(14)}$ & 0.757 & 0.708  & 1.134& 51.2             & 4.973          & 5.131 & 7.130   &8.94&  --  & G6 V  & K5 V  & SB2   \\

vB~57$^{(15)}$  & 0.491 & 0.466  & 0.530& 51.2             & 2.914          & 3.542 & 3.989   &6.46& F7 V & F6 V  & F8 V  & VB/SB2   \\

vB~58$^{(16)}$  & 0.680 & 0.642  & 0.742& 46.1             & 4.211          & 4.771 & 5.269   &7.53& G6 V & G4 V  & G8 V  & VB/SB2     \\

vB~59           & 0.543 & 0.540  & 1.492& 47.9             & 4.088          & 4.094 & 9.739   &7.49& F8 V & F9 V  & M2 V  & SB     \\

vB~62           & 0.537 & 0.518  & 1.194& 51.0             & 3.842          & 3.883 & 7.443   &7.38& F8 V & F8 V  & K5.5 V& SB1    \\

vB~63           & 0.632 & 0.630  & 1.522& 46.6             & 4.718          & 4.721 & 11.074  &8.06& G5 V & G2 V  & M3 V  & VB/SB1   \\

vB~178$^{(17)}$ & 0.837 & 0.794  & 1.295& 52.7             & 5.421          & 5.541 & 7.868   &9.03& K0 V & K0 V  & K6 V  & SB?    \\

vB~69           & 0.746 & 0.714  & 1.315& 51.5             & 5.081          & 5.159 & 7.983   &8.64& G8 V & G6 V  & K7 V  &  SB1    \\

vB~75$^{(18)}$  & 0.531 & 0.481  & 0.545& 52.0             & 3.010          & 3.644 & 4.136   &6.59& F8 V & F7 V  & F9 V  & VB/SB3   \\

vB~77           & 0.502 & 0.489  & 1.281& 47.2             & 3.680          & 3.705 & 7.800   &7.05& F7 V & F8 V  & K6.5  & SB1     \\

vB~81           & 0.470 & 0.461  & 1.327& 52.7             & 3.491          & 3.507 & 8.068   &7.10& F6 V & F6 V  & K7 V  &  SB    \\

vB~181          & 1.167 & 1.122  & 1.426& 49.4             & 6.851          & 7.050 & 8.793  &10.32& K    & K5 V  & M1 V  & PhB     \\

vB~182          & 0.884 & 0.820  & 1.062& 53.3             & 5.296          & 5.661 & 6.657   &8.93& --   & K0 V  & K4 V  & SB1   \\

vA677$^{(19)}$  & 1.23  & 1.23   & --   & 43.6             & 7.833          & 7.833 & --      &11.03&K6 V & K6 V  & --    & SB1   \\

vB~91           & 0.883 & 0.819  & 1.074& 53.1             & 5.315          & 5.659 & 6.728   &8.94& K1 V & K0 V  & K4 V  & VB/SB1   \\

vB~96$^{(20)}$  & 0.841 & 0.811  & 0.881& 50.1             & 5.011          & 5.634 & 5.910   &8.51&KO IV/V& K0 V & K2 V  & VB/SB1    \\

J301$^{(21)}$   & 1.085 & 1.085  & --   & 16.7             & 7.306          & 6.789 & --      &8.42&dK5e   & K4 V &  --   & SB1     \\

vB~102          & 0.603 & 0.576  & 1.190& 44.7             & 4.288          & 4.351 & 7.411   &7.54& G1 V & G0 V  & K5.5 V& VB/SB1    \\

J304$^{(22)}$   & 1.15  & 1.109  & 1.451&(47.1)$^{\rm (a)}$& 6.805          & 6.965 & 8.963   &10.17& K5  & K5 V  & M1 V  & SB1     \\

vB~106          & 0.669 & 0.623  & 1.089& 48.6             & 4.527          & 4.666 & 6.829   &7.96& --   & G1 V  & K4 V  & SB1     \\

vB~185$^{(23)}$ & 1.1   & 1.046  & 1.299& 51.6             & 5.907          & 6.587 & 7.886   &9.47& K3 V & K3 V  & K7 V  & SB2     \\

vB~142          & 0.665 & 0.661  & 1.501& 50.3             & 4.832          & 4.832 &10.304   &8.34& --   & G4 V  & M2 V  & SB1   \\

vB~113$^{(24)}$ & 0.549 & 0.549  & --   & 41.2             & 4.186          & 4.186 & --      &7.26& --   & F9 V  & --    & SB1   \\

vB~114          & 0.723 & 0.706  & 1.435& 49.0             & 5.089          & 5.124 & 8.845   &8.54& --   & G6 V  & M1 V  & SB1   \\

J331$^{(25)}$   & 1.41  & 1.400  & 1.421&(43.3)$^{\rm (a)}$& 7.938          & 8.616 & 8.767  &11.12& --   & K8.8 V& K9.2 V& SB2   \\

vB~115          & 0.843 & 0.814  & 1.393& 50.5             & 5.574          & 5.645 & 8.567   &9.09& --   & K0 V  & M0 V  & SB1   \\

vB~117$^{(26)}$ & 1.06  & 1.036  & 1.093& 48.02            & 6.263          & 6.544 & 6.853   &9.67& K3 V & K3 V  & K3 V  & SB2   \\

vB~119          & 0.559 & 0.528  & 1.077& 44.3             & 3.888          & 3.969 & 6.745   &7.12& --   & F9 V  & K4 V  & SB1   \\

vB~120$^{(27)}$ & 0.735 & 0.712  & 0.764& 52.7             & 4.494          & 5.150 & 5.350   &7.71& --   & G6 V  & G8 V  &  VB/SB2   \\

vB~121          & 0.504 & 0.493  & 1.324& 52.1             & 3.706          & 3.726 & 8.044   &7.29& --   & F8 V  & K7 V  & SB     \\

vB~122$^{(28)}$ & 0.541 & 0.531  & 0.553& 52.0             & 3.346          & 4.009 & 4.198   &6.76&  --  & F9 V  & F9 V  & VB/SB2    \\

vB~124$^{(29)}$ & 0.497 & 0.428  & 0.783& 54.2             & 3.057          & 3.319 & 4.879   &6.27& F5/G8& F5 V  & G5 V  & SB2   \\
\hline
\end{tabular}

$^{\rm (a)}$ Distance estimated from the photometry. \\
$^{\rm (b)}$ Calculated from V and ${\rm <m-M>_{\rm binary}}$=3.40\\
$^{(1)}$  ADS~2451. The photometry include the visual companion. Mason et al. (1993) estimated $\Delta$V=6.0$^{\rm m}$.\\
$^{(2)}$ Distance and spectral types from Strassmeier et al 1993. (B-V)$_{\rm p}$ was assigned according the
espectral type.\\
$^{(3)}$ $\Delta$V=1.75$^{\rm m}$ from Griffin \& Gunn (1981). Probable triple system.\\
$^{(4)}$ $\Delta$V=0.5$^{\rm m}$ from Griffin \& Gunn (1981).\\
$^{(5)}$ (B-V)$_{\rm p}$ measured directly during the eclipse (Schiller \& Milone 1987).\\
$^{(6)}$ $\Delta$V=0.35$^{\rm m}$ from Batten \& Wallerstein (1973). It could have an additional component.\\
$^{(7)}$ $\Delta$V=1.12$^{\rm m}$ from Griffin et al. (1982).\\
$^{(8)}$ ADS~3135. $\Delta$V=1.0$^{\rm m}$ from Wickes (1975). V=\{7.24+8.24\} from Peterson \& Solenski (1987). Visual companion included in the
 photometry.\\
$^{(9)}$ Boesgaard \& Tripicco (1986) obtained $\Delta$V=0.056$^{\rm m}$ from the ratio between spectral lines.\\
$^{(10)}$ Visual component included in the photometry.\\
$^{(11)}$ ADS~3169. Visual component included in the photometry.\\
$^{(12)}$ The contribution from the secondary is apparently negligible.\\
$^{(13)}$ The contribution from the secondary is apparently negligible.\\
$^{(14)}$ $\Delta$V=2.0$^{\rm m}$ from Griffin et al. (1985). \\
$^{(15)}$ $\Delta$V=0.45$^{\rm m}$ from Peterson \& Solenski (1987). V=\{7.01+7.46\}. The visual companion is included in the photometry.\\  
$^{(16)}$ ADS~3210. $\Delta$V=0.50$^{\rm m}$ from Heintz (1969). V=\{8.06+8.56\} from Peterson \& Solenski (1987). The visual companion is included in the 
photometry.\\
$^{(17)}$ V$_{\rm obs}$=9.03 from Mermilliod (1976).\\
$^{(18)}$ ADS~3248. $\Delta$V=0.65$^{\rm m}$ from Dombrowski (1991). Triple system. The inner pair is the SB2.\\
\indent The whole deconvolution is \{F7 V + K0 V\} + F9 V, V = \{3.644 + 5.655\} + 4.136, (B-V) = \{0.481 + 0.819\} + 0.545\\ 
$^{(19)}$ The contribution from the secondary is apparently negligible.\\
$^{(20)}$ The visual companion is {\it not} included in the photometry.\\
$^{(21)}$ Strassmeier et al. (1993) give M$_{\rm V}=6.8^{\rm m}$, D=16.7 pc and dK5e. Apparent magnitude from Eggen \& Greenstein (1965).
Distance from Borgman \& Lippicott (1983). Another possibility for the deconvolution could be
(B-V)$_{\rm calc}$=1.085, (B-V)$_{\rm p}$=1.068, (B-V)$_{\rm s}$=1.7, M$_{\rm calc}$=6.648, M$_{\rm p}$=6.687, M$_{\rm s}$=10.291,
(1+$\alpha$)=1.036, $\Delta$V=3.6$^{\rm m}$.\\ 
$^{(22)}$ $\Delta$V=2.0, Griffin et al. (1985).\\  
$^{(23)}$ The deconvolution was taken from Griffin et al. (1985). \\
$^{(24)}$ The contribution from the secondary  is apparently  negligible.\\
$^{(25)}$ $\Delta$V=0.14$^{\rm m}$, Spectral types, distance and deconvolution from Griffin et al. (1985).\\ 
$^{(26)}$ V$_{\rm primary}$=10.28, V$_{\rm secondary}$=10.59, Griffin \& Gunn (1978). Distance from Schwan (1991), Spectral types from Strassmeier et al. 
(1993)\\
$^{(27)}$ $\Delta$V=0.2$^{\rm m}$ from Thorburn et al. (1993) from the ratio between Ca{\sc I} 6718\AA. V$_{\rm obs}$=7.34, 7.59, 7.71, 7.73 and 8.18, from
Schwan (1991), Mermilliod (1976), Stern (1994) and Mason et al. (1993), respectively.\\
$^{(28)}$ ADS~3475. Also (B-V)$_{\rm obs}$=0.50 from Mermilliod (1976). $\Delta$V=0.19$^{\rm m}$ from Dombrowski (1991). The visual component is 
included in the photometry. \\
$^{(29)}$ ADS~3483. Triple system (Griffin et al. 1985):\\
\indent \{F5 V + G8 V\} + G4 V, V = \{3.4 + 5.58\} + 4.96, (B-V) = \{0.42 + 0.81\} + 0.68\\

\end{table*}

\normalsize

\section{The Deconvolution of the photometry of binary systems}
                    
\subsection{Looking for a Main Sequence}

If a binary system composed by two main sequence stars 
is unresolved, the integrated light of the system 
will be both brighter and redder than the primary star alone.
This phenomenon can be appreciated in Fig. 2a,
where we show V
vs. (B--V)$_{\rm obs}$ for all known single members of the Hyades cluster  in
the range 
$0.20^{\rm m} \le$ (B--V) $\le 1.45^{\rm m}$ (cross symbols) 
and binaries (solid and empty
circle symbols).
 The
Zero Age Main Sequence (ZAMS) from Schmidt--Kaler (1981) has been
superimposed as a 
dotted line by using a distance modulus of 3.40$^{\rm m}$
(Schwan 1991).
Since the color excess E$_{\rm (B-V)}^{\rm Hyades}$=0.00 (Gilroy
1989),
 we do not distinguish  between unreddened and reddened
colors.


 \begin{figure*}
 \vspace{9cm}
 \caption[ ]{ Color--Magnitude diagrams for all known members of
the Hyades cluster  in the range 
0.20 $\le$ (B--V) $\le$ 1.45. Single stars are shown as cross symbols
and binary systems as circle symbols (the solid circles represent the
binaries studied here and  empty circles any other binary).
(a) V against (B-V). (b) M$_{\rm V}$ against (B-V). Binary systems
are on average  $\sim$0.5$^{\rm m}$ brighter than
single stars.
  The ZAMS from Schmidt--Kaler (1981)
has been superposed 
as a  dotted line. A distance modulus of 3.4
was used in the first case. Our own MS for the
Hyades cluster is shown as a solid line (see text for details).
 }
 \end{figure*}

     On average, binary systems are more than $\sim$ 0.5$^{\rm
m}$ brighter than
single stars. In fact, a significant number of binaries are
0.75$^{\rm m}$ 
brighter than the ZAMS, which is the maximum increment in
brightness 
induced by the presence of a companion (the extreme case
of 
two stars of equal luminosities and colors). 
However, there is some scatter due to different distances,
since
the distances for the Hyades members  range from 16.7 to 72.0
pc.
These extremes correspond to J301 (Borgman \& Lippincott, 1983) 
and vB~132 (Stern, 1994). 

     Figure 2b shows M$_{\rm V}$ vs. (B-V)$_{\rm obs}$. In principal,
 the scatter introduced by the different distances
 has been removed in this figure (symbols are as in Fig. 2a). 
Only 3 binaries are above the 
limit of 0.7526$^{\rm m}$ (vB~75, vB~124 and BD+23$\degr$635). 
In fact, vB~75 and vB~124 are triple systems
(Dombrowski, 1991; Griffin et al., 1985) and   Griffin \& Gunn
(1981) found that 
BD+23$\degr$635 shows in its spectrum
a feature that cannot be attributed to the secondary, since
it does not have
periodic radial velocity shifts. They conclude that it could be
due to a third companion, 
with a long orbital period.

Based on Fig. 2b, we have catalogued vB~182, which is one of the
reddest star in our sample,  as a binary system. It is
 $\approx 0.5^{\rm m}$brighter than the ZAMS, consistent with
being a 
photometric binary (see discussion section) and inconsistent with
it being a single Hyades member assuming plausible photometric errors.

     Our deconvolution process requires a well defined MS
with  accurate photometric data. Since there is no better MS
 for a particular cluster than that computed from its members, 
we have built our own MS based on the stars belonging to the
Hyades which
have known distances (distances from Stern 1994). Because  these
distances 
were calculated based on the proper motions of each
star, they provide accurate absolute magnitudes, removing the scatter
which appears
in the V vs. (B--V)$_{\rm obs}$ diagram. In addition, this MS does not
contain any scatter 
due to evolutionary effects or due to different metallicities of
the stars 
(e.g. see ZAMS by Schmidt-Kaler, shown as dotted line in Fig.
2b). As an illustration of the effect metallicity can have, a 
  1 M$_{\sun}$ star at the age of the
 Hyades cluster has M$_{\rm V}=4.383^{\rm m}$ and (B-V)=0.411$^{\rm m}$ for  a
metallicity
Z=0.001 and M$_{\rm V}=5.159^{\rm m}$ and (B-V)=0.672$^{\rm m}$ for for a
metallicity
Z=0.020 (Schaller et al., 1992; Schaerer et al., 1993)

     To select our MS, we have traced the lower envelope by hand
on the 
M$_{\rm V}$--(B--V)$_{\rm obs}$ diagram . Since these data are extremely
accurate and they contain 
small errors either in the photometry --$\sigma$(V)=0.006$^{\rm
m}$, 
$\sigma$(B--V)=0.005$^{\rm m}$ for V $<$ 9.5-- or in the
distances for each member
--see the discussion about the absolute magnitude errors by Schwan
(1991)--, we have not considered these errors when determining the MS. The
final result 
 is shown as a 
solid line in Fig. 2b. The differences with the Schmidt--Kaler
ZAMS can be
 appreciate (dotted line). Although they are not large
($\Delta$M$_{\rm max}\sim0.3^{\rm m}$),
these differences are extremely important when the combined
photometry of a binary
 is calculated from its components.

\subsection{The deconvolution of the photometry}

We have performed a deconvolution of the photometry in order  
to obtain the true color of the primaries of the binaries,
 together with the contribution to the continuum of the secondary
component. The difference  between
 the true color for the primary 
and the observed color for the system can be up to
 0.07$^{\rm m}$, and the
 contribution to the continuum up to 50\%. In principle, it is
possible to
 reproduce the observed color and magnitude of a binary knowing
its position on
any color--absolute magnitude diagram, by adding the colors and
magnitudes of
its components. This solution is unique.
It is therefore possible to reverse the process and determine the
colors and magnitudes of the components based on the integrated
light of the binary.
Equation 1 and Eq. 2 give the  color and magnitude of 
the primary component as a function of the observed absolute magnitude
and color, as well as the magnitude and color of the secondary component:

\begin{equation}
{\rm M_V^p = - 2.5~Log~ \left\{
10^{-M_V^{obs}/2.5} - 10^{-M_V^s/2.5} \right\}   } 
\end{equation}

\vspace{5mm}
\noindent (B-V)$_{\rm p}$=
  \begin{equation}
$${\rm -2.5~Log~ \left\{
{10^{-[(B-V)_{obs}+M_V^{obs}]/2.5} - 
10^{-[(B-V)_s+M_V^s]/2.5} 
\over  10^{-M_V^{obs}/2.5} - 10^{-M_V^s/2.5}  } \right\}   }$$
\end{equation}

     Seventeen binaries have an observed difference between the
magnitudes of the components
of the system --$\Delta$m--. In these cases, we have used this
information to perform the 
deconvolution, matching the combined color and absolute
magnitude. In some cases, there are available models for the
photometry for each component, which were based  on
the study of the radial velocity traces of the spectral lines.
We only used the differences between the magnitudes, since these
models were calculated  using different Main Sequences. Therefore,
their results are slightly different from our own deconvolution.

     The results are shown in Tab. 1: Column 1 contains either
the vB (van 
Bueren, 1952), or DB (Bonner Durchmusterung), or  J (Johnson et
al., 1962)
or vA (van Altena, 1969) numbers. 
Columns 2, 3 and  4  the (B--V) color indices for the  observed
system, and the colors for
the primary and the 
secondary components, respectively. Column 5 lists the distance.
Columns 6, 7 and 8 show the absolute magnitude as observed, and as 
 calculated for the primary
and the  secondary components, whereas column 9
lists the apparent visual magnitude and columns  10, 11 and 12
contain the observed spectral type of  the binary, and the 
spectral types  assigned to the
 primary and to the secondary components based on their
calculated (B--V), following
Schmidt--Kaler (1981) or assigned by other studies.
 Column 13 gives information about the binary type.

     In order to check our deconvolution method, we have performed
the same 
procedure with other colors --namely, (V--R)$_{\rm K}$, (R--I)$_{\rm K}$
 and (V--I)$_{\rm K}$--  when available. The 
agreement was good taking into account  the uncertainties.
Each CM diagram --M$_{\rm V}$ vs. (B--V), M$_{\rm V}$ vs.  (V--R)$_{\rm K}$,
 etc-- allows us to obtain a continuum correction factor  (1$+\alpha$)
--see Section 4.2--
and the color index of the  primary for each system. 
This last quantity was transformed to
(B--V) color index by using  different relations between 
these colors --(V--R)$_{\rm K}$, (R--I)$_{\rm K}$ and (V--I)$_{\rm K}$--, 
 and (B--V) for the  single stars belonging to the Hyades Cluster.
The  dispersion for the continuum correction factor	
is 0.062 (a variation of  $\pm$6.2\% in the continuum level),
 and the  dispersion for the 
(B--V) color index is 0.015$^{\rm m}$. This last value corresponds essentially 
to the scatter introduced by the transformation between the colors.

\begin{table*}
\caption[ ]{Observational data for the late spectral type Hyades binaries. The EW \\
are not corrected for the contribution to the continuum level by the companion.}
\begin{tabular}{lrrccccc}
\hline\
Name          &  V    & SNR & EW(FeI6678) & EW(CaI6718) &EW(Fe+Li) &   EW(Li)   & Comments \\
              &       &     &  (m\AA)     &  (m\AA)     & (m\AA)   &   (m\AA)   &          \\
 (1)          & (2)   & (3) &    (4)      &    (5)      &   (6)    &    (7)     &   (8)    \\
\hline
V471 Tau      &  9.71 & 140 &    223.2    &   183.7     & 228.9   &    218.8   & (a)      \\
vA677         & 11.03 & 110 &    282.1    &   291.3     &  24.9   &     10.7   & (b)      \\
J301          &  8.42 & 132 &    247.7    &   260.3     &  32.1   &     16.2   &          \\
J304          & 10.17 &  85 &    199.5    &   205.5     &  16.5   & $\le$5.0   &          \\
vB185(p)      &  9.47 & 115 &    162.1    &   145.4     &  16.0   & $\le$2.3   &          \\
vB185(s)      &  --   &  -- &    109.1    &   119.6     &   4.6   & $\le$1.2   &          \\
J331(p)       & 11.12 &  45 &    119.2    &   133.0     &   --    & $\le$5.0   &          \\
J331(s)       &  --   &  -- &     87.2    &   110.9     &   --    &     11.3   & (c)      \\
vB117(p)      &  9.67 & 125 &    145.8    &   195.3     &   8.8   & $\le$2.4   &          \\
vB117(s)      &  --   &  -- &    114.8    &   166.4     &   6.9   & $\le$1.1   &          \\
\hline pwd

\end{tabular}
{\footnotesize

\vspace{2mm} 
 
(a) $<$EW(FeI6707$>_{\rm (B-V)=0.85}$=13.8 m\AA.\\
(b) $<$EW(FeI6707$>_{\rm (B-V)=1.23}$=14.7 m\AA.\\
(c)  EW[Li(s)+Fe(p)]$^{\rm measured}$=24.3. Assumed EW[Fe(p)]$^{\rm corrected}$=EW[Fe(s)]$^{\rm corrected}$.
EW[Fe(s)]$^{\rm measured}$=11.4\\
(p) Primary component.\\
(s) Secondary component.\\

}

\end{table*}

\section{The new Li abundances }

\subsection{The effective temperatures}

The  deconvolution process (see Section 3.2) allows us
to 
derive a more accurate (B--V) color index of the primary
component, and thus,
to calculate a more accurate effective temperature.
We have chosen a temperature scale based on the (B--V)
color. In particular,
we have used the temperature scale given by Thorburn et al.
(1993), in order to have
the maximum consistency with that work (most of
the data about 
EW(Li) were taken from their table 1). This temperature scale was
computed from 
(V--K) vs. T$_{\rm eff}$ and (B--V) vs. (V--K) relations provide
by Carney (1983), and
the Cayrel et al. (1985) zero point shift. The adopted expression
is:

\begin{equation}
{\rm T_{\rm eff} = \frac{5040}{ 0.5247 + 0.5396 (B-V)}}
\end{equation}  

The comparison between the effective temperature  derived from
the integrated color
of the binary (column 2 of Tab. 3) and from the calculated color
of the primary
(column 3 of the same table) shows that the differences are small 
(an average primary is  $\sim$150 K
hotter than the `integrated star'). However, since the
determination of Li abundances 
is extremely sensitive to the temperature, this effect is more
important in general than the
 continuum correction (see next subsection).

\footnotesize

\begin{table*}
\caption[ ]{Original abundances and corrections. The observed EW(Li) were taken primarily from
Thorburn et al. (1993). The abundances commented with SCG were calculated
by using the curves of growth by Soderblom et al. (1993). The values 
$\Delta$ N(Li) are the differences
between the abundances of the binaries and the average abundances of single stars at the same color.}
\begin{tabular}{lcccccrrrcl}
\hline\
Name& T$_{\rm eff}$&T$_{\rm eff}$& CCF       &  W(Li) &  W(Li) 
& Log N$_{\rm Li}$& Log N$_{\rm Li}$&$\Delta$ N$_{\rm Li}$& P$_{\rm orb}$ & Comments \\
    &    (K)      &Corr.       &           &Observed&Corr.   &                 &  Correct.       &            &   (d)         &          \\
 (1)&   (2)       & (3)        & (4)       &  (5)   &  (6)   &   (7)           &    (8)          &   (9)      &   (10)        &  (11)    \\
\hline
    vB~1  &  6071  &  6099  &    1.018  &    103               &    104.9  &     3.07  &     3.13  &     0.13    & --         &                \\    
V471~Tau  & 5125   &  4752  &    1.017  &    215               & 218.7     &     2.66  &     2.20  &$\ge$2.48&0.52&SGC, Log N$_{\rm Li}^{\rm Pallavic}$=2.72  \\
    vB~9  &  5558  &   --   &    --     & $\le$4               &      --   &$\le$0.80  &     --    &$\le$-0.89   &$\ge$2555                &    \\    
   vB~14  &  7036  &  7063  &    1.007  &$\le$67$^{\rm (B1)}$  &$\le$67.5  &     3.28  &     --    &$\sim$0.06   & --                      & Gap\\    
BD+23$^\circ$635&4528&4634  &    1.199  &     11               &     13.2  &     0.11  &     0.34  &$\ge$0.76&2.39&SGC, Log N$_{\rm Li}^{\rm Pallavic}$=0.26\\
  vB~162  &  5568  &  5700  &    1.724  &     41               &59$^{\rm (T)}$&  1.96  &     2.31  &     0.13    & 55.1                    &    \\    
   vB~22  &  5357  &  5542  &    1.373  &     49               &     67.3  &     1.87  &     2.28  &     0.62    & 5.61                    &    \\    
   vB~23  &  5656  &  5742  &    1.724  &     44               &     75.9  &     2.08  &     2.52  &     0.21    & 75.648                  &    \\    
BD+22$^\circ$669&4759&4902  &    1.356  &    --                &23$^{\rm (T)}$&   --   &     0.95  &$\ge$1.01&1.89&SCG, Log N$_{\rm Li}^{\rm Pallavic}$=0.89\\
   vB~29  &  6091  &  6259  &    1.398  &     21               &     29.4  &     2.03  &     2.39  &    -0.68    & 32594.91                &    \\    
   vB~34  &  6534  &  6553  &    1.950  & $\le$7.9$^{\rm (B2)}$&$\le$15.4  &$\sim$2.05 &$\sim$2.34 &$\sim$-0.08  & 3.08                    & Gap\\    
   vB~38  &  7227  &  7364  &    1.066  &$\le$19.6$^{\rm (B1)}$&$\le$20.9  &     2.89  &     --    &$\sim$-0.39  & 2.14                    & Gap\\    
   vB~40  &  6083  &  6225  &    1.092  &    112               &    122.3  &     3.17  &     3.46  &     0.39    & 4.00                    &    \\    
   vB~39  &  5659  &  5659  &$\sim$1.000&     44               &     44.0  &     2.08  &     2.08  &     0.03    &$\ge$2555                &    \\    
   vB~42  &  5394  &  5422  &    1.016  &     11               &     11.2  &     1.10  &     1.14  &    -0.10    & --                      &    \\    
   vB~50  &  5936  &  6024  &    1.048  &     73$^{\rm (R)}$   &     76.5  &     2.62  &     2.74  &    -0.18    & 12045                   &    \\    
   vB~52  &  5951  &  5951  &$\sim$1.000&     85               &     85.0  &     2.77  &     2.77  &    -0.04    & --                      &    \\    
  vB~140  &  5400  &  5558  &    1.159  & $\le$3               & $\le$3.5  &$\le$0.52  &$\le$0.74  &$\le$-0.98   & 156.4                   &    \\    
   vB~57  &  6382  &  6594  &    1.661  &     39               &     64.8  &     2.69&$\approx$3.18&$\sim$0.43   & 2292                    & Gap\\    
   vB~58  &  5652  &  5785  &    1.631  &     74               &107$^{\rm (T)}$& 2.43  &     2.87  &     0.34    & 10106                   &    \\    
   vB~59  &  6163  &  6176  &    1.005  &     82               &     82.4  &     2.98  &     3.00  &    -0.06    &$\ge$2555                &    \\    
   vB~62  &  6188  &  6267  &    1.038  &    126               &    130.8  &     3.47  &     3.64  &     0.57    & 8.88                    &    \\    
   vB~63  &  5821  &  5829  &    1.003  &     70               &     70.2  &     2.51  &     2.52  &    -0.02    & 2595.3                  &    \\    
  vB~178  &  5162  &  5288  &    1.117  &      5               &      5.6  &     0.49  &     0.68  &    -0.09    & --                      &    \\    
   vB~69  &  5435  &  5539  &    1.074  &     11               &     11.8  &     1.15  &     1.28  &    -0.37    & 41.6                    &    \\    
   vB~75  &  6212  &  6427  &    1.793  &     49               &     87.9  &     2.66  &     3.35  &     0.28    & 21.25                   &    \\    
   vB~77  &  6335  &  6391  &    1.023  &     29$^{\rm (R)}$   &     29.7  &     2.47  &     2.54  &$\sim$-0.45  & 238.9                   & Gap\\    
   vB~81  &  6475  &  6516  &    1.015  &$\le$17.6$^{\rm (B2)}$&$\le$17.9  &$\sim$2.39 &$\sim$2.42 &$\sim$-0.37  & --                      & Gap\\    
  vB~181  &  4365  &  4460  &    1.201  &$\le$30$^{\rm (D)}$   &$\le$36.0  &$\le$0.35  &$\le$0.57  &     --      &(11.92)$^{\rm (P_{rot})}$& SCG\\    
  vB~182  &  5031  &  5211  &    1.399  &     10               &     14.0  &     0.67  &     1.02  &     0.48    & 358.4                   &    \\    
   vA677  &  4241  &  4240  &$\sim$1.000&     10.7             &     10.7  &    -0.30  &    -0.30  &$\ge$0.38    & 1.49                    & SCG\\  
   vB~91  &  5034  &  5213  &    1.373  & $\le$4               & $\le$5.5  &$\le$0.26  &$\le$0.59  &$\le$0.05    & 9131                    &    \\    
   vB~96  &  5150  &  5237  &    1.775  &      6               &     10.7  &     0.57  &     0.92  &     0.31    & 4748                    &    \\    
    J301  &  4540  &  4540  &$\sim$1.000&  16.2                &     16.2  &     0.31  &     0.31  &$\ge$0.81&1.788&SCG, Log N$_{\rm Li}^{\rm Pallavic}$=0.23\\
  vB~102  &  5928  &  6032  &    1.060  &     85               &     90.1  &     2.75  &     2.88  &    -0.05    & 10592                   &    \\    
    J304  &  4400  &  4488  &    1.159  & $\le$5               & $\le$5.8  &$\le$-0.43 &$\le$-0.24 &     --      & 60.821                  & SCG\\  
  vB~106  &  5690  &  5854  &    1.136  &     73               &     82.9  &     2.56  &     2.68  &     0.08    &$\approx$3652            &    \\    
 vB~185(p)&  4529  &  4627  &    1.302  & $\le$2.3             & $\le$3.0  &$\le$-0.59 &$\le$-0.38 &  -- &276.76&SGC, Log N$_{\rm Li}^{\rm Pallavic}\le$-0.43\\
 vB~185(s)&   --   &  4112  &    4.311  &$\le1.2$              & $\le$5.2  &     --    &$\le$-0.77 &     --      & 276.76                  & SGC\\  
  vB~142  &  5704  &  5718  &    1.006  &     54               &     54.3  &     2.26  &     2.27  &     0.04    & --                      &    \\    
  vB~113  &  6139  &  6139  &$\sim$1.000&     77               &     77.0  &     2.88  &     2.88  &    -0.16    &$\approx$2555            &    \\    
  vB~114  &  5509  &  5565  &    1.032  &     29               &     29.9  &     1.71  &     1.77  &     0.03    &$\ge$1825                &    \\    
   J331(p)&  3921  &  3937  &    1.871  &$\le$5.0              & $\le$9.3  &$\le$-0.89 &$\le$-0.60 &   --        &8.495                    & SGC\\
   J331(s)&   --   &  3921  &    2.148  &     11.3             &     24.3  &    -0.51  &    -0.16  &$\ge$0.00    &8.495                    & SGC\\
  vB~115  &  5145  &  5228  &    1.068  &      3               &      3.2  &     0.25  &     0.37  &    -0.21    &$\approx$1460            &    \\    
 vB~117(p)&  4596  &  4650  &    1.751  & $\le$2.4             & $\le$4.2  &$\le$-0.48 &$\le$-0.15 &   -- &11.93&SGC, Log N$_{\rm Li}^{\rm Pallavic}\le$-0.25\\
 vB~117(s)&   --   &  4522  &    2.330  & $\le$1.1             & $\le$2.6  &     --    &$\le$-0.55 &     --      & 11.93                   & SGC\\  
  vB~119  &  6099  &  6225  &    1.077  &     71               &     76.5  &     2.77  &     2.98  &    -0.09    & --                      &    \\    
  vB~120  &  5470  &  5545  &    1.831  &     28               &     51.3  &     1.65  &     2.10  &     0.43    & --                      &    \\    
  vB~121  &  6326  &  6374  &    1.019  &$\le$116.9$^{\rm (B2)}$&$\le$119.0&$\sim$3.54 &$\sim$3.62 &$\sim$0.62   & 5.75                    & Gap\\    
  vB~122  &  6171  &  6212  &    1.840  &     46               &     84.6  &     2.58  &     3.06  &    -0.01    & 16.30                   &    \\    
  vB~124  &  6356  &  6670  &    1.372  &$\le$11$^{\rm (B2)}$  &$\le$15.1  &$\sim$2.03 &$\sim$2.34 &$\sim$1.48   & 143.53                  & Gap\\    
\hline 
\end{tabular}
{\footnotesize

\vspace{2mm} 
 $(1+\alpha) = 1+10^{\rm-[(V_s-V_p)/2.5]}$, correction of continuum for primary components.
 
$^{\rm (B)}$ Boesgaard \& Tripicco (1986a). The Li EW includes  the weak FeI 6707.4 \AA:  \\  
\indent $\,\,\,\,\,$ $^{\rm B1}$ Abundances from that paper. $^{\rm B2}$ W(Fe+Li)$_{\rm BT}$ = W(Fe+Li)$_{\rm T}$ -- 2.00 m\AA.     \\  
$^{\rm (D)}$ Duncan \& Jones (1983).  \\  
$^{\rm (R)}$ Rebolo \& Beckman (1988).  W(Li)$_{\rm RB}$ = 1.3$\times$W(Li)$_{\rm T}$ -- 19.00 m\AA.      \\  
$^{\rm (T)}$ Corrected by Thornburn et al. (1993).      \\  
$^{\rm (P_{rot})}$ We have classified vB~181 as a photometric binary, and assumed that the orbital period is equal to the
photometric period.\\

 }
\end{table*}

\normalsize

\subsection{The calculation of the Li  equivalent widths  for
binaries}

     The measured EW have been corrected for the contribution
to the continuum by the companion. The formula used for the
correction was:

\begin{equation}
{\rm EW_{corrected} = EW_{measured} \times (1+\alpha)}
\end{equation}

\noindent where,

\begin{equation}
{\rm \alpha = 10^{-\frac{\rm (V_s-V_p)}{2.5}}, \, primary \,
components}.
\end{equation}

\begin{equation}
{\rm \alpha = 10^{+\frac{\rm (V_s-V_p)}{2.5}}, \, secondary \,
components}.
\end{equation}

We have compared the results obtained by using Eq. 4 with those
obtained
by using a more sophisticated expression --e.g., see Boesgaard \&
Tripicco (1986b), 
which takes into account the effective temperatures and  radii
to
compute the contribution of each component to the local continuum
around 6700 \AA. The differences are quite small, with the
advantage
for the simpler expression that it does not need to introduce
additional
expressions between the radium and any other known stellar
parameters of the 
components, such as (B--V).

     The calculated continuum correction factors (1+$\alpha$) are listed in
column
4 of Tab. 3. As can be seen, there is a correspondence between
(1+$\alpha$)
and the type of spectroscopic binary, when available (SB2, those
stars which
have spectral lines arising from both components, and SB1, where
only the 
primary is seen). 
A high value of (1+$\alpha$) means that both components can be
seen in the
 spectrum and, in fact, they correspond to SB2 binaries.
Batten (1973) has suggested that a secondary  component 
1.75$^{\rm m}$ fainter than the primary [i.e. (1+$\alpha$)=1.2] 
should not be observed in the
spectrum. Our data generally confirm this rule.

     Original EW(Li)s are listed in column 5 of Tab. 3, whereas
 the corrected EW(Li)s are in column 6. The six EWs taken from Boesgaard
\& Tripicco
(1986a) include the weak Fe{\sc I} 6707.4 \AA\, feature.
However, all are mid--F dwarf and this contribution is negligible
($\sim$ 4 m\AA). The EWs with the B1 superindex
correspond to the 
original value of Boesgaard \& Tripicco (1986a), and we also
have conserved their Li abundance value.
Those stars with the B2 superindex --Boesgaard \& Tripicco
(1986a)-- and with the R superindex
--Rebolo \& Beckman (1988)--  were corrected following Thorburn
et al. (1993)
 in order to avoid systematic effects.

\subsection{The correction of the Li  abundances  for binaries}

 We have calculated  final abundances by using two different
sets of curves of growth.
The first set was taken from Pallavicini et al. (1987) and 
the T$_{\rm eff}$ range is 6500--4500 K. The input model
atmospheres were
those of Gustafsson et al. (1975), Bell et al. (1976) and Kurucz
(1979). 
The second set of curves of growth was taken from Soderblom et
al. (1993a)
 and it contains curves of growth down to 4000 K. This
calculation  used the 
model atmospheres of Bell et al. (1990).
 Both sets were calculated in LTE conditions.
In the linear part of the curves of growth (the part that we have
essentially used),
 the average difference between both sets is $\sim$0.07 dex, with
higher abundances in
the case of Soderblom's et al. (1993a) curves.

Recently, Carlsson et al. (1994) 
have obtained an empirical  relation to correct the deviation of the
Li{\sc I}
6707.8 \AA{ } curves of growth in LTE from the NLTE. Since our
goal is to perform a comparison between
stars having the same metallicity and color, we did not apply
this correction.

There are several reasons why we have used simultaneously 2 sets
of curves of growths: First, we used the
 Pallavicini et al. (1987) curves because they reproduce 
 fairly  well   the abundances obtained
by Thorburn et al. (1993). However, we needed curves of growth at
lower temperatures
 to obtain abundances for stars redder than  (B--V)=0.9$^{\rm m}$.
 We could have used the differences between the  curves at  T$_{\rm
eff}$=4500 K to correct those by Soderblom
et al. (1993a) at lower temperatures, following Duncan \& Jones
(1983). 
However, we preferred not to make this correction because our conclusions
 would be unaffected by the shift and there is no reason to believe 
such a zero point shift would 
result in better abundances for the cooler stars.
 In fact, we studied binaries redder and bluer than (B-V)=0.9$^{\rm m}$
with two different goals:
We have performed a comparative study between single
stars and binary systems
of the same color for those stars having (B--V)$<$0.9$^{\rm m}$
(and in this case we use the Pallavicini et al. curves of growth),
whereas the aim of the study made with the subsample 
at lower temperature was to detect the presence of the Li doublet
in close binaries and not in wide
binaries (we use for this subsample the Soderblom et al. curves of growth).

 Columns 7 and 8 of Tab. 3 show the uncorrected and corrected
abundances for our sample of binary stars, respectively
--where Log N(Li) = 12+Log ${\rm
\left\{\frac{N(Li)}{N(H)}\right\} }$.
The first one was 
calculated by using the uncorrected T$_{\rm eff}$ for the
binary and the measured
EW(Li) --columns 2 and 5 of Tab. 3, respectively. The second one
was computed by using the 
assigned T$_{\rm eff}$ to the primary component and the
corrected EW(Li)
--columns 3  and 6 of Tab. 3, respectively.

\subsection{The error estimation}

In order to estimate the final errors of our Li abundances, we
should consider the different
 sources of the uncertainties, namely, the error in the
measurement process of the
EWs, the errors in the effective temperature scale, those given
by the deconvolution calculation 
and, finally, those characteristic of the curves of growth.

The EWs analyzed here were taken primarily 
from Thorburn et al. (1993) or
are original from this work.
In the first case, the typical SNR is $\sim$200. Since most of
these systems do not have
 a high value of v~sin~i and  are not very cool, the
determination of the
 continuum is quite clear, which allows one to obtain EWs with
small
internal errors. Thorburn et
al. (1993) give $\sigma_{\rm EW}$=2.0 m\AA.

In the case of the EWs measured directly by us (V471~Tau, vA~677,
J301, J304, vB~185, J331 and vB~117),
the errors are sometimes more important. There are different
reasons for this fact: The SNR is
worse in these spectra ($<$SNR$>\sim$100). Due to their lower
effective temperature, new lines
and molecular bands could appear close to the Li{\sc I} 6707.8
\AA{ } doublet (however, in all
cases we attributed the feature only to the Fe+Li blend). Finally,
some of the systems have very
broad lines in their spectra, because they are very rapid
rotators. These phenomena lead to
 uncertainties in the continuum determination and blendings with
other lines.
We estimate the measured EW(Li) errors for V471~Tau as 25 m\AA,
for vA~677 as 6 m\AA{ }, for 
J301 as 2 m\AA{ }  and for the secondary of J331 as 8.2 m\AA.
 For the other binaries, we were only able to  obtain upper
limits.

As a quite conservative estimate, we assign an error of 100 K to 
the effective  temperature. 

The estimation of the errors of the deconvolution process is more
complicated. A shift of
+0.1$^{\rm m}$ in the locus of the Hyades MS (an estimation by
eye of 
the error in in the location of the MS), causes differences of
-0.01$^{\rm m}$ in the (B--V)
color index and 0.03$^{\rm m}$ in the absolute magnitude for the
primaries 
calculated with the deconvolution process (they are hotter and
fainter). 
This shift adds another $\sim$30 K to the error in the
temperature
and  5\% the continuum correction.

In total, the average error in our abundance determination
(without the error of the curves of growth)
can be estimated as 0.25 dex.

\section{Results}

As shown  in Tab. 1 and Tab. 3, we have been able to correct
the colors,
temperatures, EW(Li)s and Li abundances of the primary components
(and
in some cases the secondaries) of binary
systems due to the presence of a companion. Although the
correction for 
T$_{\rm eff}$ and EW(Li) are very small (${\rm <\Delta
T_{eff}>}$=150 K and
${\rm <\Delta  EW(Li)>}$=8 m\AA), these differences are enough to
increase the
 abundances significantly for a number of stars.

     Although our corrections for double-lined spectroscopic
binaries
 are essentially equal to
those
performed by Thorburn et al. (1993), and they are based in the
same raw 
data (most of them have published photometry for the components
or at least
 magnitude differences), there are also differences  (see next
section). 
It is important to note that they did not perform any correction
for the single--lined spectroscopic 
binaries because they found a tight correlation between the
temperatures
obtained from (B-V) and the total EWs of several lines of Fe{\sc
I}
and
 Al{\sc I}. This fact allowed them to define an independent
temperature
 scale (T$_{\rm spect}$). Since the differences between both
temperatures were
very small, they pointed out that it was not necessary to correct
the abundances either due to reddening of the colors or the
continuum 
contribution due to the presence of the secondary component.
However, the combination  of both effects is important enough 
to change substantially the estimated Li abundances (see Tab. 3)
and the correction must be done to avoid biasing the Li abundances of the
single--lined spectroscopic binaries.


 \begin{figure*}
 \vspace{9cm}
 \caption[ ]{  EW(Li) versus (B--V).  
Single stars are shown as
crosses and binaries are shown as open circles (without the
deconvolution) and solid circles (with the deconvolution)
(a) Plot for the uncorrected EW(Li) data.
(b) Plot for corrected EW(Li).
Note that the very short--period TLBS V471 Tau is out of the frame in both cases. 
 }
 \end{figure*}

\section{Discussion}

     Figure 3a shows EW(Li) versus (B--V) for the uncorrected
data. Crosses represent 
single stars and open circles the uncorrected data for the
binaries. Despite 
 the fact that binaries follow the same trend as single stars
(the EWs decreases
with the (B--V), and there is a clear gap for the mid--F
stars), there are also 
differences: There are several cool stars with large EW(Li)
 --BD+23$\degr$635, BD+22$\degr$669, V471~Tau, vA~677, J301 and
J331.
For stars in this spectral range, Li is rapidly destroyed during
the premain-sequence phase (PMS) or the first stages of the MS
evolution, as
 has been shown by different studies in $\alpha$~Per, Pleiades
and UMa~ Group
(Boesgaard et al. 1988a,b; Soderblom et al. 1993a,b;
Garc\'{\i}a--L\'opez et al. 1994).
 The first two systems were studied by Thorburn et al. (1993).
They interpreted their
 overabundances as an effect of their rapid rotation, since these
systems are 
tidally locked binary systems --TLBS-- (P$_{\rm orb}$=2.39 and
1.89 d, respectively). The 
angular momentum transfer from the binary orbit to  stellar 
rotation, which takes
 place in TLBS due to tidal effects (Zahn
1994 and references therein),  inhibits  the
Li depletion
(Pinsonneault et al. 1989; Zahn 1994). This occurs because
there is less radial differential rotation, and thus
less mixing between  the interior and the convective envelope. 
The last four systems in the list above are also tidally locked
(P$_{\rm
orb}$=0.52, 1.49, 1.788 and 8.495 d,
respectively), in support of this model. 
In fact, their high EW(Li)s and 
abundances are really remarkable. The same situation appears 
for vB~22, vB~62 and vB~121, which have overabundances and short
rotational periods
(P$_{\rm orb}$=5.61, 8.88 and 5.75 d, respectively). 
There are several alternatives to this model for the Li depletion.
 Garc\'{\i}a--L\'opez
\& Spruit (1991) and Montalban \& Schatzman (1994) have shown that internal
gravity waves reproduce the Li depletion in F and G--K spectral type stars, 
respectively. On the other hand, Spruit (1987) has
suggested that dynamo--induced magnetic fields, which are more 
intense in rapid rotators like close binaries, can inhibit the turbulent
mixing. The combination of both mechanisms could also explain the 
overabundance we have found in TLBS.  
Moreover, rapid rotation can change the distribution of temperature in
the stellar interior (Mart\'{\i}n \& Claret 1995) and, therefore, the 
Li depletion.  
However, all these scenarios 
(mixing due to radial differential rotation, gravity waves, etc)
do not fit with 
vB~40, which has P$_{\rm orb}$=4.00 d and an {\it uncorrected}
EW(Li) similar to those of single
stars at the same (B--V).

     Also, there is an apparent  shift in  the
position of the
Li gap for the binaries relative to that for the single stars.
The minimum EW(Li)
appears at
(B--V)$_{\rm single}\approx0.43^{\rm m}$ and (B--V)$_{\rm
binary}\approx0.50^{\rm m}$. Is this a 
characteristic due to the binarity {\it per se} or it is just a
bias introduced
 by the combined photometry of both stars in the binary
systems?

     Figure 3b is as Fig. 3a, except that we have plotted the
corrected EW(Li)
and (B--V) --solid circles-- instead of 
the measured EW(Li) . The
excesses for
the TLBS still appear, but  vB~40  
{\it has an excess now}.
Moreover, the possible shift of the Li gap  has disappeared,
adding support for
 the validity of our method of deconvolution.

     The different behavior of Li between 
binaries and the single stars are more conspicuous in  Fig.
4, which shows the corrected data for 
the Hyades stars in a  Log N$_{\rm Li}$--T$_{\rm eff}$
plane. The average Li abundance for single stars at each
effective  temperature 
is shown as a solid line and binaries are shown as solid symbols
(triangles show TLBS and circles other binaries). 
For these corrected data, there are six notable ways in which
the single and binary stars differ:


 \begin{figure*}
 \vspace{9cm}
 \caption[ ]{   Log N(Li) against T$_{\rm eff}$. The corrected abundances 
and effective temperatures are shown.  The average Li abundance for
single stars are represented as a solid line, 
 TLBS are shown as solid triangles, other binaries belonging to
the Hyades are shown as solid circles. TLBS (open triangles) 
and other stars (open circles)
belonging to the Pleiades are also included.

 }
 \end{figure*}

\begin{itemize}

\item All TLBS  with T$_{\rm eff}$ cooler than 6500 K have overabundance
(BD+23$\degr$635, BD+22$\degr$669, V471~Tau, vB~22, vB~34, vB~40,
vB~62, vA~677, J301, J331  and vB~121). Note that we have not corrected
the abundance of vB~38, since its high  effective temperature is out of
the temperature ranges of both sets of curves of growth.

\item Some longer orbital  period binaries also have abundances less than
the average Li abundance of single stars (solid line)
 at the same T$_{\rm eff}$.

\item There are some binaries with unknown or longer orbital period larger than 
that of TLBS and Li overabundances
(vB~58, vB~182, vB~96 and vB~120). In this case, it is not possible to relate
these high abundances with  rapid rotation due to tidal forces.
 
\item The gap for mid--F stars is centered at similar
temperature, but the average
 abundance could be  higher for binaries than for single stars.

\item The maximum abundance of  binary stars are 0.4 dex higher
than that of single
 stars.

\item The larger scatter on the abundances is still present (note
the large differences
between the abundances for those systems with T$_{\rm
eff}\approx5600$ K and also for
 those  at  T$_{\rm eff}\approx6300$ K).

\end{itemize}

Similar behavior has been found by Deliyannis et al. (1994) in a
TLBS belonging to the 
older cluster  M67. They found that this system has a Li abundance
a factor 2 or more 
larger than any other M67 single stars, and a factor 3 larger
than the average Li peak
 region.

Figure 4 also includes  the Pleiades
data. TLBS belonging to the Pleiades are shown 
as empty triangles and other Pleiades stars (single or binary) as empty circles.
 The Pleiades is
younger than the Hyades (8.00$\times$$10^7$ yr, Meynet et al.
1993) and their 
stars with 5500 K $\ge$ T$_{\rm eff}$ $\ge$ 4000 K still have
high Li surface abundances.
 As can be seen, the Li abundances in the cool Hyades TLBS are
similar to that of the 
younger single stars. Thus, the Li depletion must have been
inhibited 
during the PMS and/or the first part of the MS
evolution, when a single star loses its angular momentum at a
very high rate
(Stauffer et al. 1994). As can be seen, the two TLBS in the
Pleiades do not have 
a Li overabundance, as expected (Soderblom et al. 1993a; Ryan \&
Deliyannis 1994). 
This  supports the idea that most of the 
Li depletion inhibition occurs during the MS
life--time, rather
than during the  premain sequence phase. The data  for these two
systems were calculated using the corrected color and  EW(Li),
 following the method described in Section 3.

In order to clarify the role of the binarity on the Li
phenomenon, we have calculated
the differences --$\Delta$ N(Li)-- between the abundances of 
the binaries and the average abundance
 of single stars at the same color (solid line in Fig. 4) for
the Hyades stars. The average
 behavior for single stars was calculated by fitting a third
order polynomial for (B--V)$\ge0.5^{\rm m}$. This curve was extended beyond
(B--V)=0.9$^{\rm m}$ (T${\rm eff}\le5000$ K) by using the the upper
limit of the abundance of those systems which are not tidally locked
(dashed line).


 \begin{figure*}
 \vspace{9cm}
 \caption[ ]{  Differences between the corrected Li abundance
for each binary and the average Li abundance for single stars 
with the same temperature plotted versus the orbital period 
 in the range T$_{\rm eff}$ $>$ 6500 K. 
 All binaries having P$_{\rm orb}\le9$ d  have significant
overabundances.
 }
 \end{figure*}

Fig. 5 shows this difference
against the orbital period for the binaries in the range
${\rm T_{eff}} \le$ 6500 K. It demonstrates clearly that
 all binaries having P$_{\rm orb}\le9$ d  have significant
overabundances. Zahn 
(1989) pointed out that binaries having P$_{\rm orb}\la8$ d
should be synchronized 
before their arrival on the MS.
During almost the whole  synchronization process and afterward,
the orbital angular
momentum behaves as a source of angular momentum for the rotation,
replacing the angular 
momentum loss (Soderblom 1983).
 This allows the TLBS to avoid the differential rotation between 
the core and the convective envelope
(MacGregor \& Brenner 1991) 
and the mixing mechanism involved which is the result of it
(Pinsonneault et al. 1989).
 Thus, their Li depletion during the
PMS phase would be inhibited. In fact, if the inhibition process
would take place essentially
during the PMS phase, some very short binaries   would
retain the original Li abundance of the interstellar cloud from
which the Hyades
cluster would be eventually formed.

Binaries with periods $\ga$ 8-9 d are not expected to arrive on
the MS with their rotational periods synchronized to their orbital
periods.  However, binaries with orbital periods just slightly longer than
this cutoff (P$\sim$10-20 d) should exchange some orbital angular momentum
during PMS evolution and while on the MS.   The lack of any apparent
lithium enhancement in the one Hyades G dwarf binary system in this period
range (vB 122, P$\sim$16.3 d) suggests that synchronization is required
to prevent PMS lithium depletion, and not just ``some" angular momentum
transfer.  It is also true, however, that some older TLBS with orbital
periods larger than P$_{\rm cutoff}$ of the Hyades do have lithium enhancements
(Barrado et al. 1994).  This suggests that lithium depletion also normally
occurs on the MS and that tidal synchronization even then can prevent
or diminish further MS lithium depletion.

     VB~181 deserves special attention. This star has not been
previously classified as 
binary. However, its position in the CM diagram (it is
0.5$^{\rm m}$ brighter
in absolute magnitude than a MS star of
 the same color) and its slightly faster 
than  average rotation compared to
 stars of the same color (Radick et al. 1987),
suggest that  it could be a binary star 
(our deconvolution gives K5 V + M1 V spectral types for both
components). This star
has a very high upper limit for its abundance (EW(Li)$_{\rm
corrected}\le36.1$ m\AA, 
whereas EW(Li)$_{\rm single}\sim$10 m\AA{ } for
(B--V)$\sim$0.80). 
There is another observation of this star, by
Zapala (1972),
who reported the presence of Li. However, Duncan \& Jones (1983)
only found an upper 
limit and specificly  did not confirm the presence of Lithium.
The binary system vB~117 also  has a similar orbital period (11.93 d).
We were also only  able to get an upper limit to the abundance in this
case.
Since these  stars could have orbital periods
 in the border between the 
synchronization during the PMS  and the MS phases
(we have assumed that vB~181 has an orbital period synchronized with the
rotational period, P$_{\rm rot}$=11.92 d), more accurate spectral
 observations should be made to improve their Li abundance estimates.

The most conspicuous case of Li overabundance in TLBS in the Hyades
is V471~Tau. This
is a system composed of a K2~V star and a white dwarf
(Ruci\'nsky 1981). The orbital period 
is the shortest in the sample (P$_{\rm orb}$=0.51118 d, Young
et al. 1983).
 Because of our lack of knowledge of the 
 original masses, rotation and orbital periods, and the
interaction 
mechanisms that  have taken place before and during the
evolution of the former
more massive star (the white dwarf) to this stage, it is only
possible to
speculate about the reason of this overabundance.
As well as the Li
depletion inhibition explained above (the very short rotational
and orbital periods would have prevented any core--envelope
mixing and thus the surface would have retained its primordial
Li), Li could be created by
spallation reactions
 in the envelope of the giant star which eventually evolved to
white dwarf. This 
material could be transported to the now K2~V star via the overflow 
of the Roche limiting surface, following a similar mechanism of
that one postulated for
 Walter (1995) has also  pointed out that
it has very cool structures on its surface. These structures could also
change the Li equivalent width.
 Barium stars. It is also possible that our original EW(Li) has
 a larger than average systematic
error, since the rotational broadening will blend other lines
with Li{\sc I} 6707.8 \AA.

     Three other systems are worthy of discussion. VB~9, vB~29 and vB~140 have
upper limits 
for their abundances much lower than those expected  due to their
temperature. The simplest
explanation would be that they are not real members of the
cluster. However, their photometry
agrees with Hyades membership as does, for vB~29 and vB~140, the
proper motion and the radial velocity
 (Schwan 1991 and references therein). Both vB~9 and vB~29 have X-ray
luminosities appropriate for  Hyades members of that color; 
vB~140 however has an X-ray upper limit that is below that of typical
Hyades members of its color (Stern et al. 1994).

     On the other hand, vB~182 and vB~58 (and possible vB~96)
have clear overabundances
despite their long orbital periods. In this case, we cannot
explain the overabundances 
by appealing to  angular momentum transfer, since their separations
are too wide to have tidal effects
(P$_{\rm orb}$=358.4 d and 27.7 yr, respectively) Are they
real members? If they are, 
which mechanism would explain their high abundances? The
photometry of the 3 systems 
 and the radial velocity and proper motion for vB~58 and vB~96
(Schwan 1991)  correspond to Hyades members. They may have been
rapid rotators at Pleiades age, and for these reason they inhibit 
the Li depletion till that moment.  After that, they have been
burning Li as single stars do, but they have not lost their 
excess yet.  This scenario suggests that they
might still be relatively rapid  rotators. Their slightly 
greater than average X--ray luminosities agree with this possibility
 (Stern 1994).

\section{Summary}

The main result of this work is the conclusion that every TLBS in
the Hyades cluster
have Li overabundance. This  inhibition
of the Li depletion would have happened essentially during the
PMS  and/or the first
stage of the MS evolution.
Further studies in tidally locked binaries in younger clusters 
would show exactly when the inhibition of the Li
depletion takes place and the scale of the phenomenon.

\acknowledgements{This research has made use of the Simbad database, 
operated at CDS, Strasbourg, France.
The comments by an anonymous referee have
been very useful.
DBN thanks the Real Colegio Complutense at 
Harvard University for the support as a Visiting Fellow and
 Charles Prosser for his help
during the data gathering.}

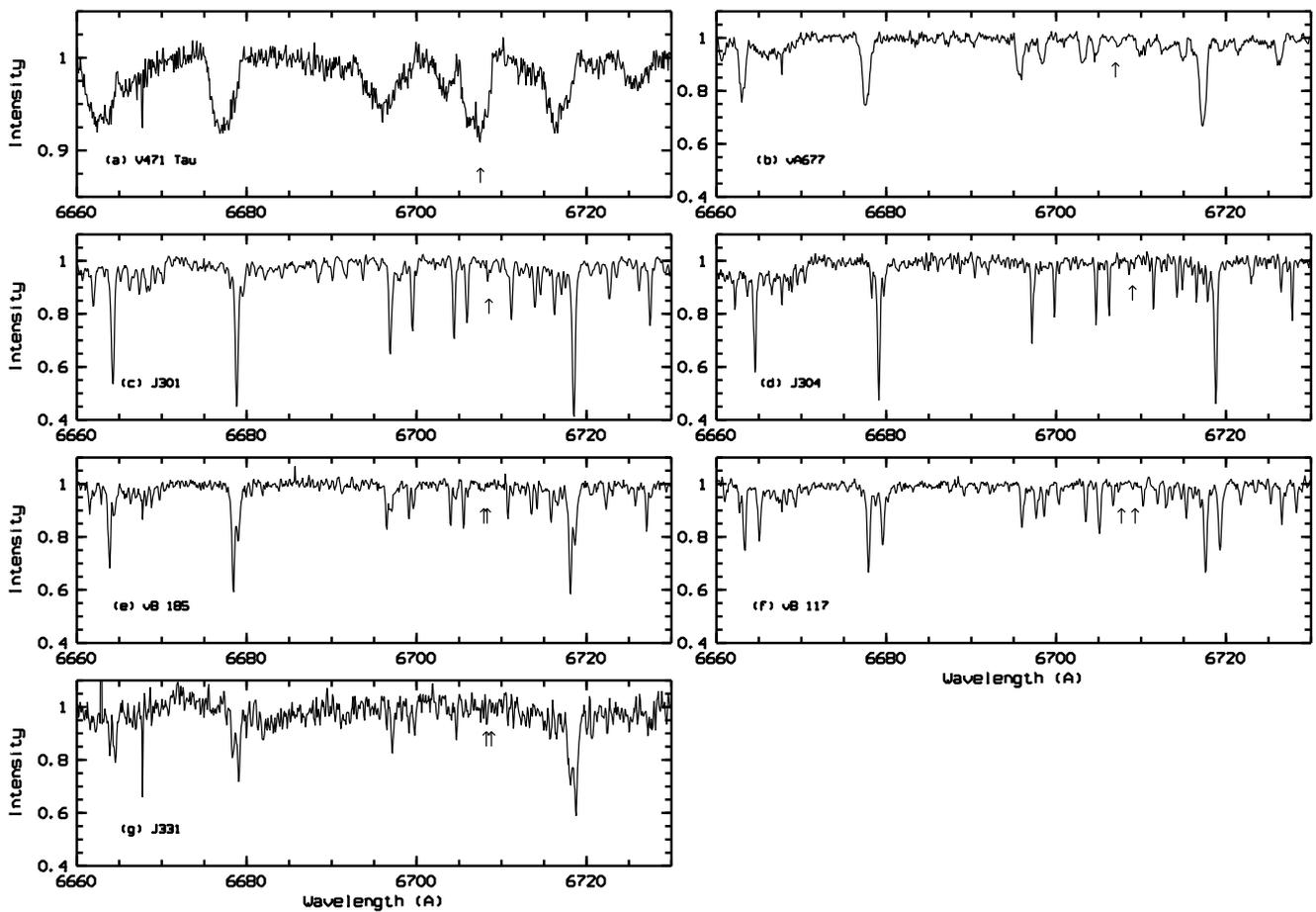
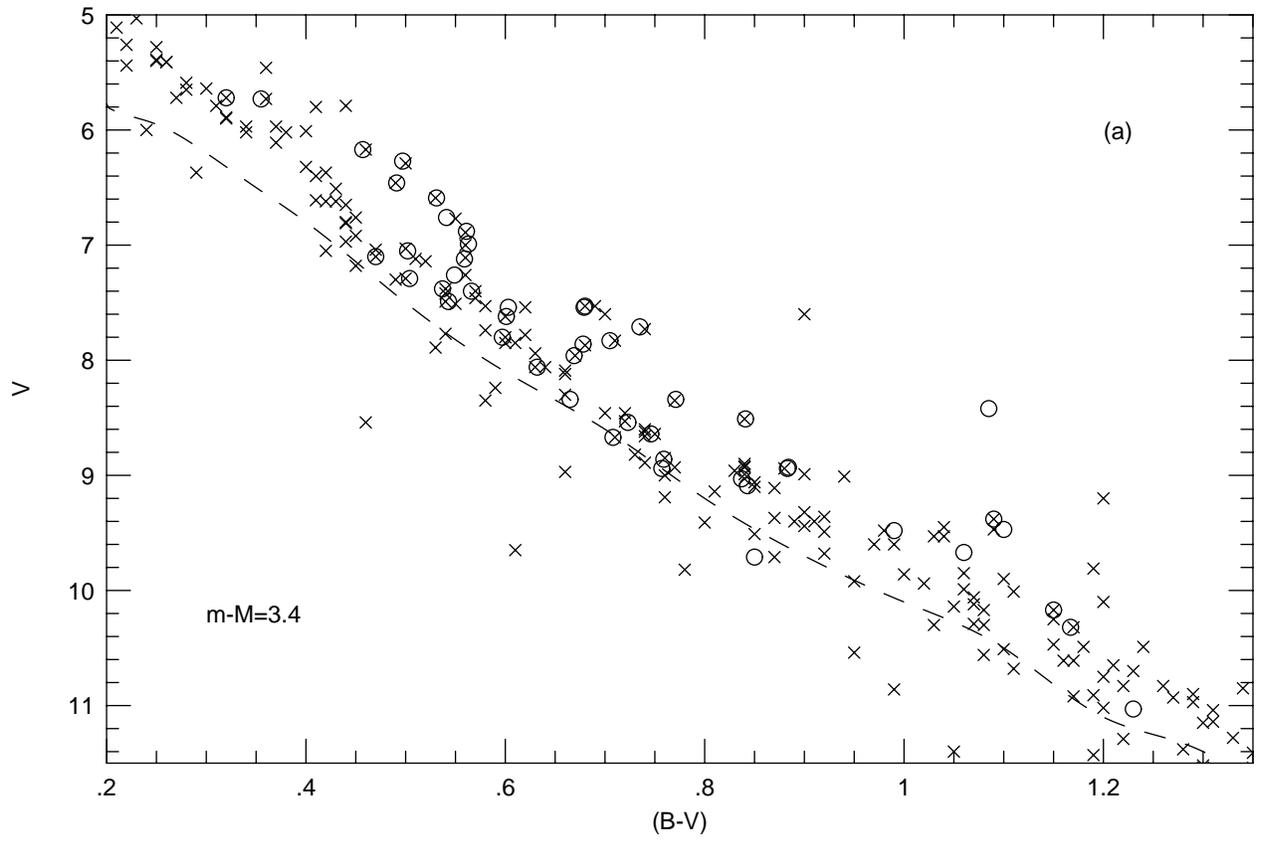
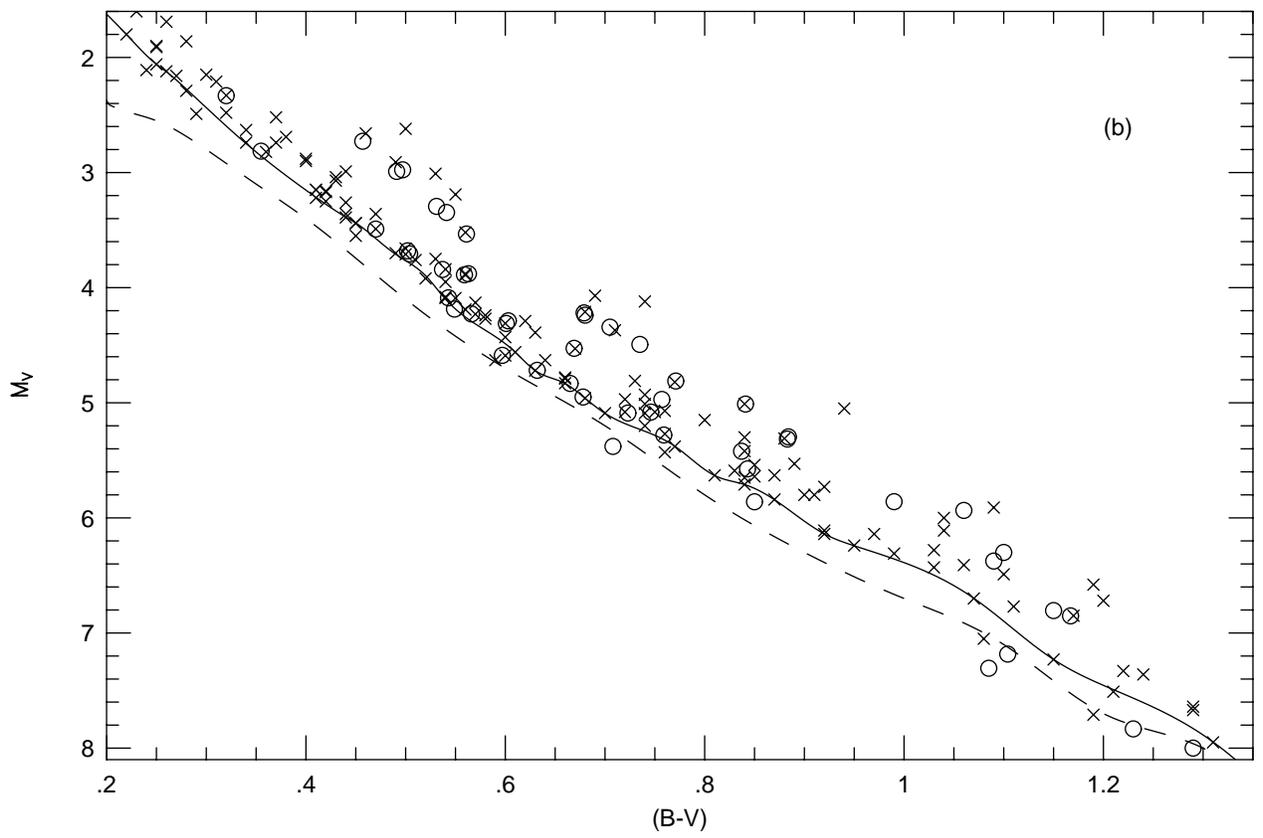
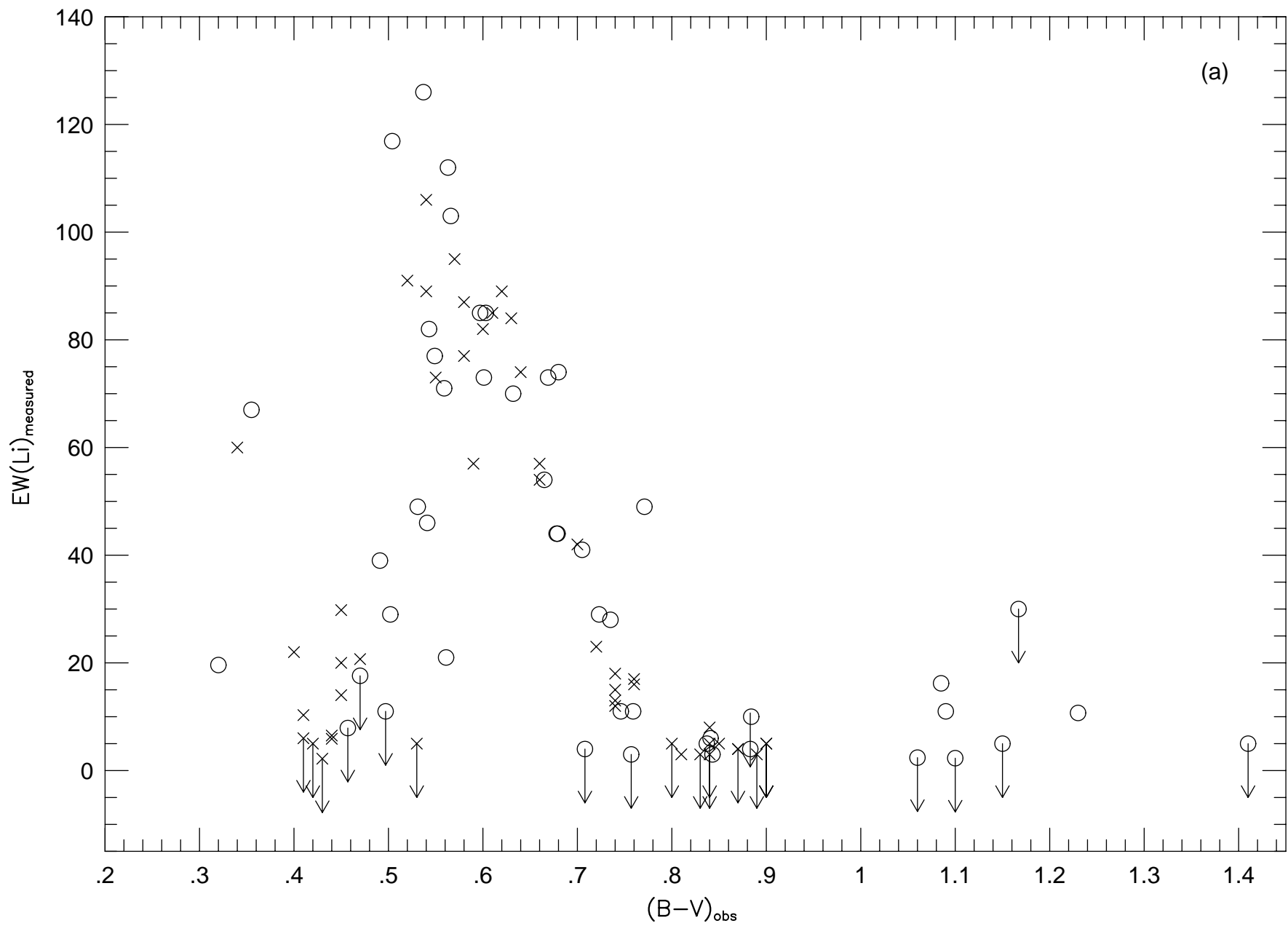
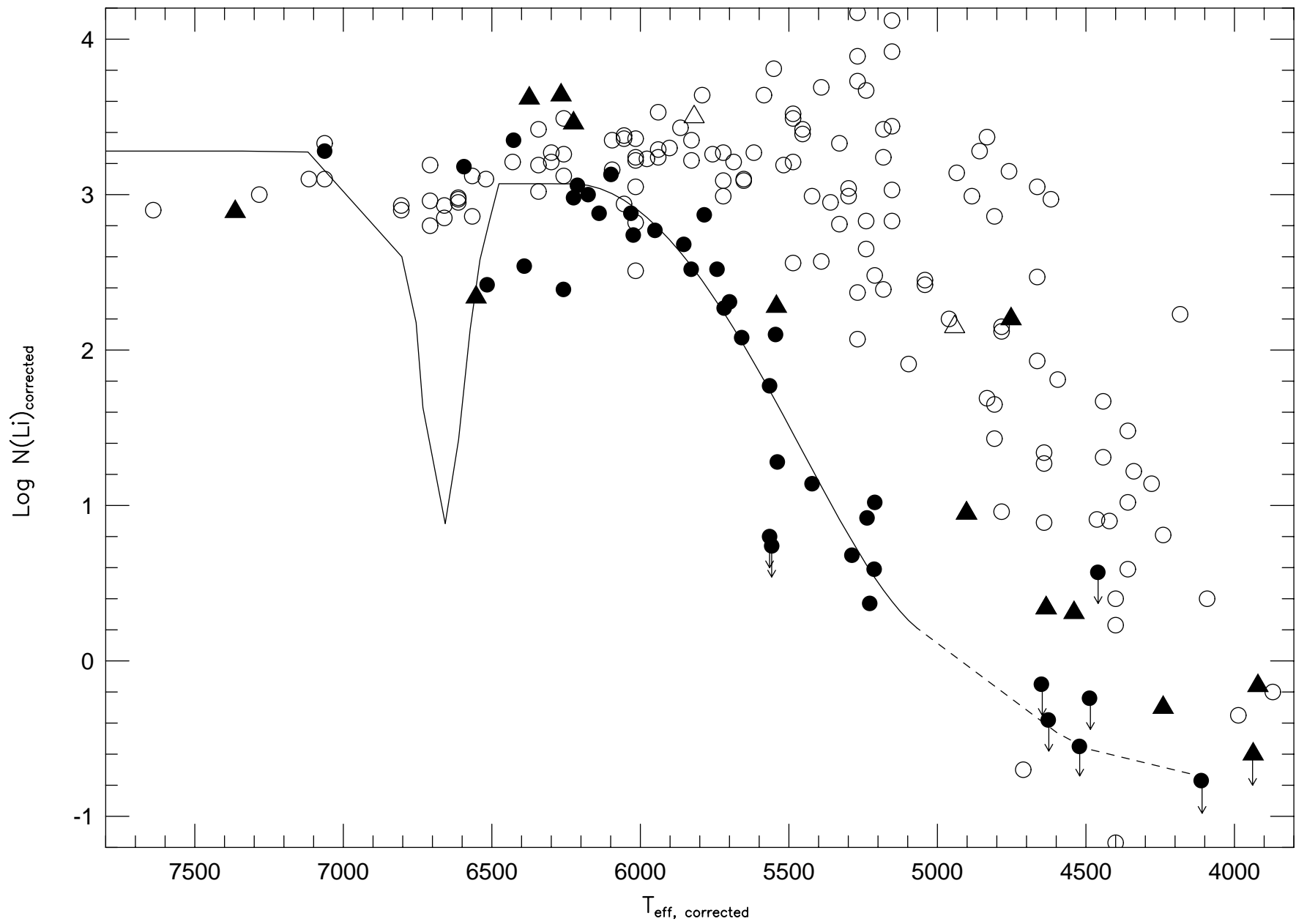
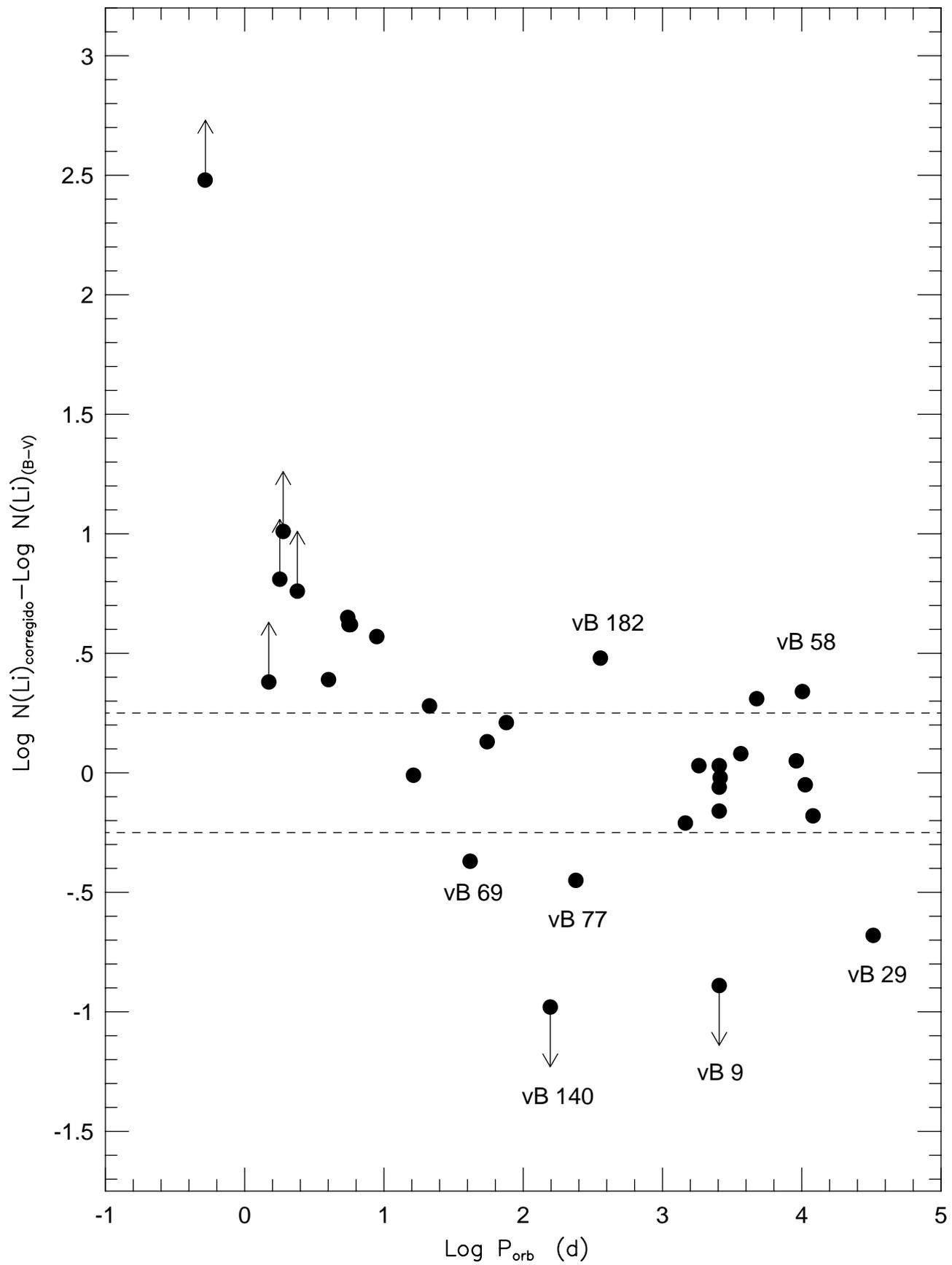
\end{document}